# Stratification Dynamics of Titan's Lakes via Methane Evaporation

Jordan K. Steckloff[1,2,3], Jason M. Soderblom[1], Kendra K. Farnsworth[4], Vincent F. Chevrier[4], Jennifer Hanley[5,6], Alejandro Soto[7], Jessica J. Groven[6,8], William M. Grundy[5,6], Logan A. Pearce[3,6,9], Stephen C. Tegler[6], Anna Engle[6]

Abstract

Saturn's moon Titan is the only extraterrestrial body known to host stable lakes and a hydrological cycle. Titan's lakes predominantly contain liquid methane, ethane, and nitrogen, with methane evaporation driving its hydrological cycle. Molecular interactions between these three species lead to non-ideal behavior that causes Titan's lakes to behave differently than Earth's lakes. Here, we numerically investigate how methane evaporation and non-ideal interactions affect the physical properties, structure, dynamics, and evolution of shallow lakes on Titan. We find that, under certain temperature regimes, methane-rich mixtures are *denser* than relatively ethane-rich mixtures. This allows methane evaporation to stratify Titan's lakes into ethane-rich upper layers and methane-rich lower layers, separated by a strong compositional gradient. At temperatures above 86K, lakes remain well-mixed and unstratified. Between 84 and 86K, lakes can stratify episodically. Below 84K, lakes permanently stratify, and develop very methane-depleted epilimnia. Despite small seasonal and diurnal deviations (<5K) from typical surface temperatures, Titan's rain-filled ephemeral lakes and "phantom lakes" may nevertheless experience significantly larger temperature fluctuations, resulting in polymictic or even meromictic stratification, which may trigger ethane ice precipitation.

[1] Massachusetts Institute of Technology, Department of Earth, Atmospheric, and Planetary Sciences, 77 Massachusetts Ave, Cambridge, MA 02139. Corresponding author: jordan1@mit.edu
[2] Planetary Science Institute, 1700 East Fort Lowell, Suite 106, Tucson, AZ 85719
[3] University of Texas at Austin, Department of Aerospace Engineering and Engineering Mechanics, Aerospace Engineering (ASE) Building 2617 Wichita Street, C0600, Austin, Texas 78712
[4] University of Arkansas, Arkansas Center for Space and Planetary Sciences, F47 Stone House North, Fayetteville, AR 72701
[5] Lowell Observatory, 1400 W Mars Hill Rd, Flagstaff, AZ 86001
[6] Northern Arizona University, Department of Astronomy and Planetary Science, Flagstaff, AZ 86011
[7] Southwest Research Institute, 1050 Walnut St. Suite 300, Boulder, CO 80302
[8] Washington State University, Institute for Shock Physics, Pullman WA 99164
[9] Steward Observatory, University of Arizona, 933 North Cherry Avenue, Tucson, AZ 85721

1. INTRODUCTION

Numerous physical mechanisms can spontaneously stratify terrestrial lakes into low-density upper layers (epilimnia) and high-density lower layers (hypolimnia). For example, summer heating may form warm epilimnia that float atop cooler hypolimnia. Similarly, salinity or other chemical gradients can cause density differences that drive stratification. Although epilimnia and hypolimnia are typically thermochemically uniform (well-mixed), the interface between them exhibits a steep gradient in temperature (thermocline), salinity (halocline), and/or chemical composition (chemocline). Certain conditions may stabilize this pycnocline (density gradient), and cause permanent stratification (a meromictic lake). Other conditions can evolve the limnia toward a uniform density, causing the lake to mix/overturn (a monomictic or dimictic lake; *Löffler 2004*).

Following methane's discovery in Titan's atmosphere (*Kuiper 1944*) and predictions of methane–ethane oceans (*Sagan & Dermott 1982; Lunine et al. 1983*), the Cassini mission discovered hydrocarbon lakes on Titan (*Stofan et al. 2007*). The composition of Titan's lakes, however, leads to complex behaviors that complicate a simple application of terrestrial limnologic principles. Whereas terrestrial lakes are composed of a single chemical species ($H_2O$) with small amounts of solutes (~1 mM or ~0.001 mole fraction; *Lehman et al. 1998*), Titan's hydrocarbon lakes (predominantly a methane-ethane mixture; *Lunine et al. 1983; Stevenson & Potter 1986; Mitri et al. 2007*) dissolve a large, temperature-dependent amount of atmospheric nitrogen (*Stevenson & Potter, 1986; Battino et al. 1984; Malaska et al. 2017; Hartwig et al.2018*). Methane, ethane, and nitrogen molecules interact with one another, causing significant deviations from ideal behavior. Previous efforts to model the thermodynamic evolution of Titan's lakes either neglect evaporation-induced changes in chemical composition and physical properties (*Mitri et al. 2007*), or use simplifying assumptions that neglect the full non-ideal interactions between all three of these chemical species (*McKay et al. 1993; Tokano et al. 2009*). Because liquid ethane is significantly denser than liquid methane, these simplifying assumptions led to a false belief that methane-rich mixtures are always buoyant upon ethane-rich mixtures, and that evaporation, which preferentially removes methane due to its higher volatility (*Stevenson et al. 1986; Luspay-Kuti et al. 2015*), will keep lakes well-mixed. Here, we develop a numerical model to investigate how non-ideal interactions affect the physical properties, structure, dynamics, and evolution of shallow lakes on Titan due to methane evaporation

2. METHODS AND RESULTS

We built the TITANPOOL numerical model to investigate the thermodynamic, physical, and chemical evolution of non-ideal methane–ethane–nitrogen lakes under conditions present at the surface of Titan. We focus on how shallow, ephemeral rain-filled lakes *(Barnes et al. 2013; Soderblom et al. 2016)* and polar "phantom lakes" (*MacKenzie et al. 2019*) evolve due to methane evaporation. Rain droplets on Titan are thought to form initially from nearly pure methane. However, modeling work has shown that when droplets fall through an atmosphere with moderate ethane relative humidity (50%), the precipitation that reaches the surface can contain equal parts methane and ethane (*Graves et al. 2008*). Thus, the initial composition of the ephemeral, rain-filled lakes we consider could plausibly have initial methane:ethane ratios as low as 50%, even prior to the onset of methane evaporation at the surface.

Rather than explicitly modelling evaporation (which cools the lake if evaporative heat flux exceeds sensible heating), TITANPOOL uses the GERG-2008 equation of state (*Kunz et al. 2007; Kunz et al. 2012*) to compute a lake's density as a function of temperature and methane-ethane-nitrogen composition. Although measured surface temperatures on Titan range between 89 and 96 K (*Cottini et al. 2012; Jennings et al. 2016*), previous work has shown that evaporative cooling and thermodynamic evolution both while falling (*Graves et al. 2008*) and on the ground (*Rafkin & Soto 2019*) can allow surface liquids to be several Kelvin cooler than the surrounding land environment. We therefore consider this full range of plausible *liquid* surface temperatures. We then use the results of our calculations to investigate how methane evaporation affects the mixture's composition, and density, which may lead to stratification. Lastly, we check our numerical results and confirm our analysis with analogous laboratory experiments.

*2.1 The TITANPOOL Numerical Code*

To investigate the thermophysical evolution of a hydrocarbon mixture on Titan, we developed the TITANPOOL numerical model in Matlab using the Reference Fluid Thermodynamic and Transport Properties Database (REFPROP). TITANPOOL is purpose-built to study the thermodynamic, chemical, and physical evolution of a non-ideal, ternary mixture of methane, ethane, and nitrogen under the conditions present at the surface of Titan. Since REFPROP cannot directly compute the equilibrium composition and material properties of a mixture, TITANPOOL contains a series of subroutines that call REFPROP to obtain the properties of a specified mixture at a specified temperature and pressure. TITANPOOL uses these functional calls to linearly interpolate the equilibrium liquid composition and its properties. REFPROP natively computes the liquid density of a ternary mixture, along with compositions in vapor-liquid equilibrium (VLE).

REFPROP is produced, updated, maintained, and benchmarked by the National Institute of Standards and Technology (NIST; *Lemmon et al. 2010*), and has already been used to study the behavior of Titan's lakes (*Hartwig et al. 2018*). REFPROP is an industry standard program that first computes the Helmholtz free energy of pure and multicomponent material mixtures to compute thermodynamic and material properties (such as density and equilibrium composition. For hydrocarbon and nitrogen mixtures, REFPROP uses the Group Européen de Recherches Gazieres (GERG-2008) equation of state (*Kunz et al. 2007; Kunz and Wagner 2012*), which is commonly used in the oil and gas industry. The GERG-2008 equation of state most accurately models the material properties of hydrocarbon and nitrogen mixtures in its "Normal Range of Validity" (90K $\leq$ T $\leq$ 450K, $p \leq$ 35MPa), for which the availability of extensive experimental data has reduced uncertainties in liquid density computations to less than 0.5% and uncertainties in the *pTxy* relationships (pressure, temperature, and gas and liquid phase compositions) to less than ~5% (*Kunz & Wagner 2012*). Outside of this range, GERG-2008 necessarily must extrapolate from the available experimental data.

In this "Extended Range of Validity" (60K $\leq$ T $\leq$ 700K, $p \leq$ 70MPa), little experimental data exist to accurately assess model uncertainties. *Kunz & Wagner (2012)* estimate that the uncertainties in liquid component density and *pTxy* relationships are expected to be no more than a few percent. Recent experimental work of methane-ethane-nitrogen mixtures found that GERG-2008 indeed has errors in the densities of such mixtures of no more than ~7% for conditions present at the surface of Titan (~1.5 bar, ~90-94 K; *Richardson et al. 2018*). Comparing GERG-2008 with our experimental work (~1.5 bar, 83 – 87K), we confirm that errors in pressure in this "Extended Range of Validity" are similarly no more than 7% (see section 2.3). However, recent experimental work on methane-ethane-nitrogen mixtures found that GERG-2008's errors became quite significant (up to ~18% in density) at pressures much higher than those at the surface of Titan or the depths of its lakes (*Richardson et al. 2018*). Nevertheless, since we are only considering shallow lakes at Titan's surface, we restrict our concern to GERG-2008's accuracy at Titan's surface pressure (~1.5 bar), which is sufficient for our purposes.

We interfaced REFPROP with Matlab to construct our model of the thermodynamic properties of a lake of methane, ethane, and nitrogen on Titan. We assume that the entire lake is at a uniform temperature (thermal equilibrium – a reasonable assumption for ponds that we anticipate to be only a few meters deep), and that no sensible heat flows into or out of the lake. Due to the low volatility of ethane relative to methane under Titan surface conditions (*Stevenson and Potter 1986; Tokano 2009*), we also assume that no ethane evaporates and that the entire mixture is in vapor-liquid equilibrium (VLE) at 1.5bar with the gas boundary layer above the surface of the lake. We use REFPROP to compute the equilibrium mole fraction of nitrogen dissolved in the liquid as a function of lake temperature and methane hydrocarbon fraction ($f_{methane}$) as a function of temperature at Titan surface conditions:

$$f_{methane} \equiv \frac{x_{methane}}{x_{methane} + x_{ethane}}$$

We then use REFPROP to compute the liquid density of the lake, also as a function of temperature and methane hydrocarbon fraction.

This model neglects the effects of hydrostatic pressure and temperature gradients/heating on stratification. Nitrogen becomes increasing soluble in methane-ethane mixtures at increasing hydrostatic pressures, and thus would drive up liquid densities significantly in the depths of deep liquid bodies (e.g., Kraken Mare); this would further stabilize lake stratification. However, these effects are negligible over the depths of the ephemeral ponds and shallow lakes that we consider here. Similarly, solar heating would tend to warm the surface layers of liquid bodies, causing epilimnia to lose dissolved nitrogen and become even more buoyant, allowing stratification to occur at even warmer temperatures. Therefore, our simple model may err conservatively on the conditions required for stratification, which places strong constraints on the dynamic behavior of Titan's lakes.

*2.2 Results of Numerical Modelling*

Nitrogen solubility in liquid alkanes depends strongly on alkane complexity, being highest in the simplest alkane, methane (*Battino et al. 1984*). It also shows strong temperature dependence, with cooler temperatures leading to increased nitrogen solubility (*Battino et al. 1984; Malaska et al. 2017; Hartwig et al. 2018*). In alkane mixtures, non-ideal interactions between alkane species strongly affects nitrogen solubility (Figure 1). Because nitrogen is considerably denser than either methane or ethane, its abundance strongly influences the density of mixtures, and can cause methane-rich mixtures to become denser than ethane-rich mixtures under certain conditions (Figure 1).

Titan's shallow, ephemeral (*Barnes et al. 2013; Soderblom et al. 2016*) and polar "phantom" (*MaKenzie et al. 2019*) lakes are likely cooler than the ~89–96K surface (*Cottini et al. 2012; Jennings et al. 2016*). Methane evaporation allows falling rain to reach the surface ~4K cooler (*Graves et al. 2008*) than the surrounding atmosphere. Recent mesoscale modelling of the interactions between air and lakes shows that methane evaporation may further cool lake temperatures by up to 5 (or more), depending on the initial conditions (*Rafkin & Soto 2019*). This level of cooling occurs when the atmosphere has less than 20% relative methane humidity and the lake is shallow (~1 meter deep; *Rafkin & Soto 2019*). The methane humidity across Titan is generally greater than 50%, and even higher near the poles (*Lora & Adámkovics 2016*). Nevertheless, a higher relative humidity would corelate with a 37.5% reduction in evaporative cooling, thus lowering the lake temperature by up to 3–4 K. Thus, the combination of these effects allows the temperatures of these shallow lakes to plausibly be as low as 83K.

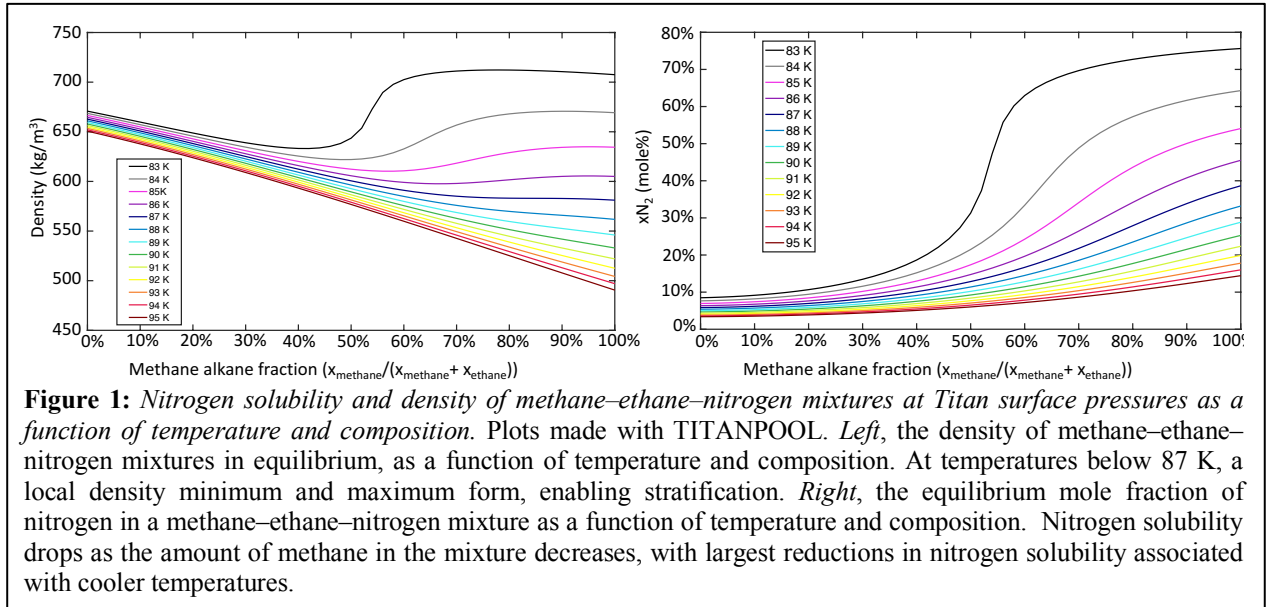

**Figure 1:** *Nitrogen solubility and density of methane–ethane–nitrogen mixtures at Titan surface pressures as a function of temperature and composition.* Plots made with TITANPOOL. *Left*, the density of methane–ethane–nitrogen mixtures in equilibrium, as a function of temperature and composition. At temperatures below 87 K, a local density minimum and maximum form, enabling stratification. *Right*, the equilibrium mole fraction of nitrogen in a methane–ethane–nitrogen mixture as a function of temperature and composition. Nitrogen solubility drops as the amount of methane in the mixture decreases, with largest reductions in nitrogen solubility associated with cooler temperatures.

We find that the density of methane–ethane–nitrogen mixtures at typical Titan surface temperatures (> 87K) decreases monotonically with increasing methane hydrocarbon fraction. However, significant deviations from this monotonic behavior appear at temperatures below 87K, producing a local density maximum and minimum at methane alkane mole fractions ($f_{methane}=\frac{x_{methane}}{x_{methane}+x_{ethane}}$) of 70–90 mole%, and 40–70 mole%, respectively. Below 84K, the methane-rich local maximum composition ($f_{methane}$ = 70–90 mole%) is denser than a methane-free, ethane–nitrogen binary mixture ($f_{methane}$ = 0%). This surprising density behavior has important implications for the evaporation-dependent evolution of Titan's lakes, the stratification of Titan's lakes, and the limnological layer from which ethane ice might precipitate.

This composition and temperature dependence allows Titan's lakes to stratify. Above 87K, evaporative methane loss enriches the lake's surface layer in ethane, producing a denser liquid regardless of composition (Figure 1). Thus, the lake's surface layer will sink into, and mix with the hypolimnion, preventing stratification. Similarly, the surface layer of cool (<87K) methane-poor lakes (methane alkane fraction less than the composition of the local density minimum, Figure 1, left) will become denser from evaporation, mixing the lake and preventing stratification.

However, methane-rich lakes (relative to the local minimum's composition) below 87K evolve differently (Figure 2). Preferential methane evaporation from the surface layer *decreases* the epilimnion's density (if also poorer in methane than the local density *maximum*), producing a buoyant epilimnion that fails to mix with the hypolimnion (Figure 2a). Further evaporation (Figure 2b) evolves the epilimnion's composition toward the local density minimum (Figure 2c), producing a relatively methane-poor epilimnion and methane-rich hypolimnion, separated by a strong chemocline. Continued evaporation (Figure 2d) increases the epilimnion's density above the local density minimum. Eventually, evaporation will drive the epilimnion's density above that of the hypolimnion, causing the lake to overturn (epilimnion sinks and mixes with the hypolimnion, Figure 2e). Evaporation from the well-mixed lake would then form a new buoyant epilimnion, and this stratification cycle would repeat.

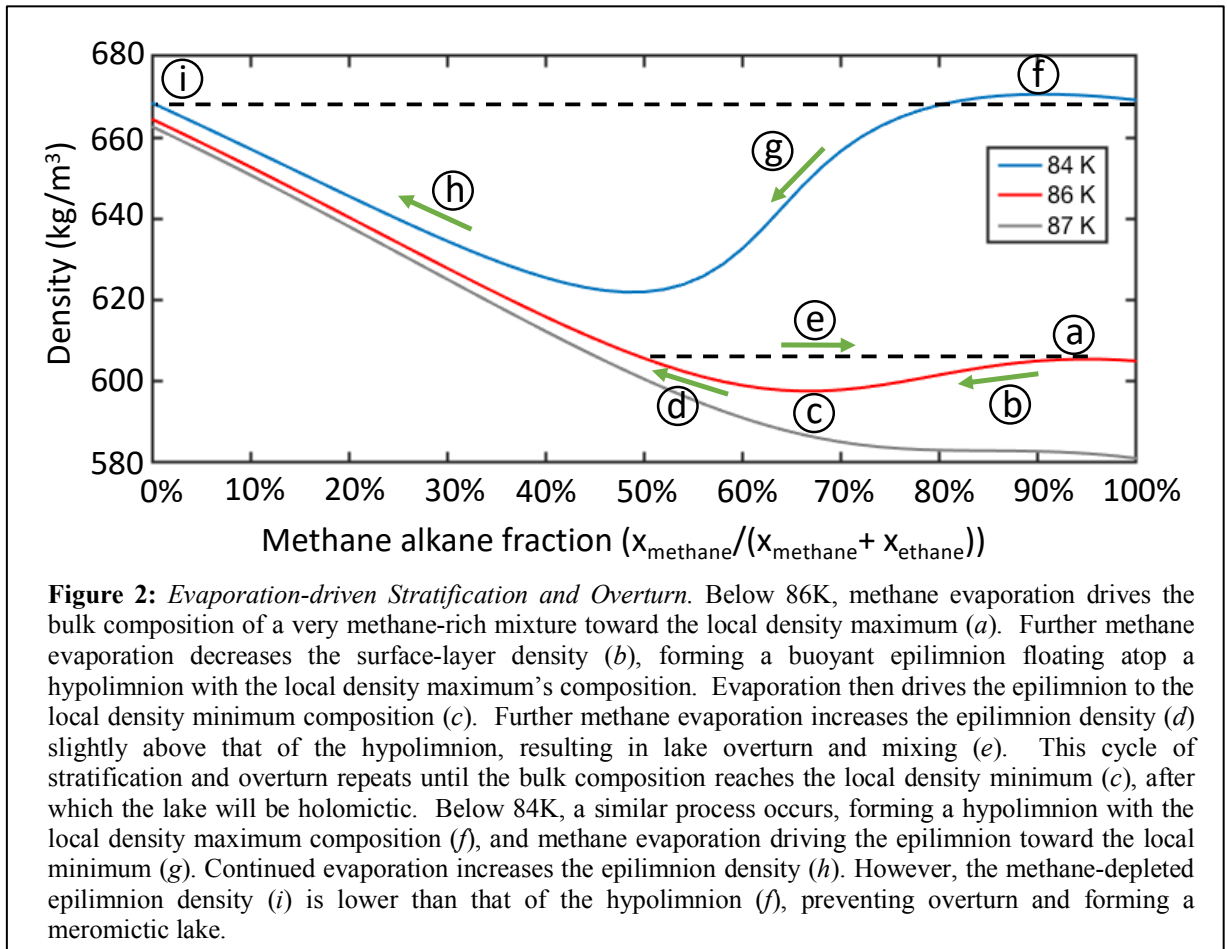

**Figure 2:** *Evaporation-driven Stratification and Overturn.* Below 86K, methane evaporation drives the bulk composition of a very methane-rich mixture toward the local density maximum (*a*). Further methane evaporation decreases the surface-layer density (*b*), forming a buoyant epilimnion floating atop a hypolimnion with the local density maximum's composition. Evaporation then drives the epilimnion to the local density minimum composition (*c*). Further methane evaporation increases the epilimnion density (*d*) slightly above that of the hypolimnion, resulting in lake overturn and mixing (*e*). This cycle of stratification and overturn repeats until the bulk composition reaches the local density minimum (*c*), after which the lake will be holomictic. Below 84K, a similar process occurs, forming a hypolimnion with the local density maximum composition (*f*), and methane evaporation driving the epilimnion toward the local minimum (*g*). Continued evaporation increases the epilimnion density (*h*). However, the methane-depleted epilimnion density (*i*) is lower than that of the hypolimnion (*f*), preventing overturn and forming a meromictic lake.

Each overturn event mixes the ethane-rich epilimnion with the methane-rich hypolimnion; the new bulk lake composition is a linear combination of the epilimnion and hypolimnion compositions. The addition of the epilimnion's ethane-rich material to the hypolimnion may result in a bulk lake that is supersaturated in nitrogen (see Figure 3). This supersaturated nitrogen would likely exsolve, forming bubbles — such bubbles have been detected experimentally under similar conditions (*Farnsworth et al. 2019; Richardson et al. 2019*). On Titan's larger seas, overturn-driven bubble formation may produce the observed "Magic Islands" (*Hofgartner et al. 2016*). However, it is unclear if these large lakes cool sufficiently to facilitate stratification-driven bubbling.

Further evaporative methane loss depletes the bulk lake of methane (composition moves leftward in Figures 1–2). Eventually, the bulk lake composition will evolve past the local density minimum, inhibiting further stratification.

Below 84K, methane evaporation drives the composition of the epilimnion toward the local density minimum (Figure 2g). Although further evaporation (Figure 2h) increases the epilimnion density, the epilimnion never sinks and never mixes with the methane-rich hypolimnion because the methane-free, ethane–nitrogen binary mixture (Figure 2f) is still less dense than the methane-rich local density maximum (Figure 2i). The lake becomes meromictic. These dynamic paths are illustrated in Figure 4.

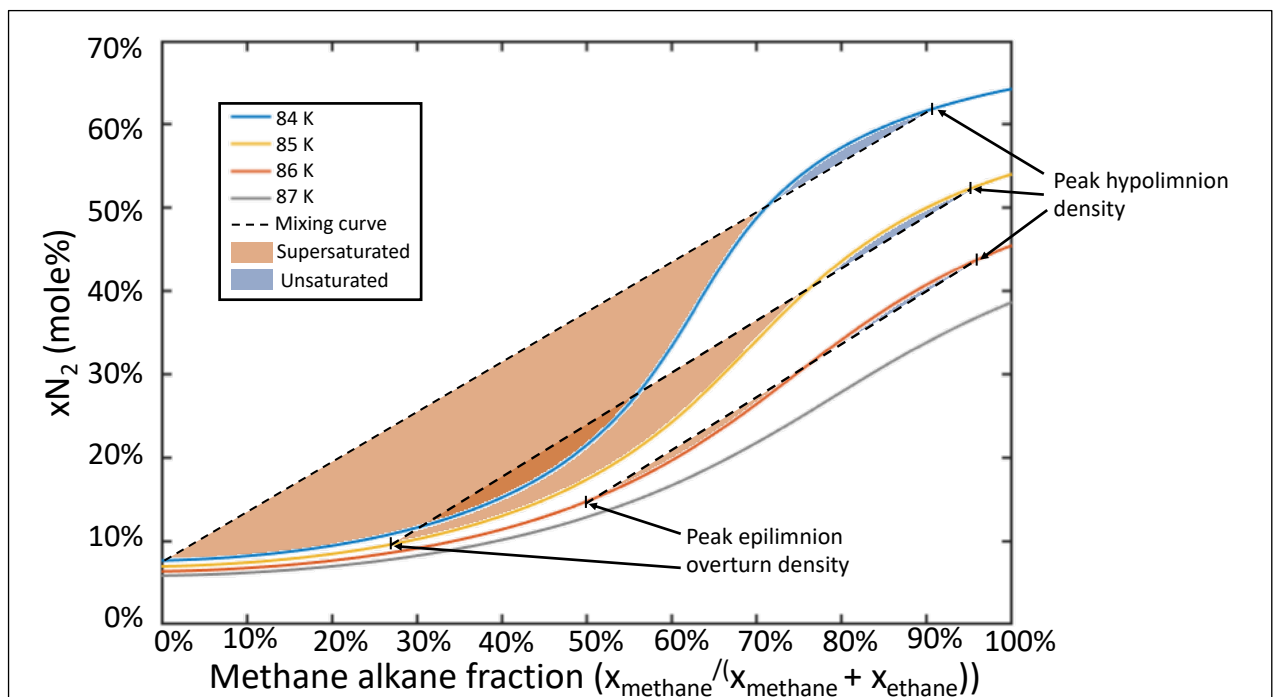

**Figure 3:** *Overturn events can lead to supersaturated or unsaturated lakes.* Upon overturning, a lake will mix the epilimnion and hypolimnion, forming a new uniform mixture. The possible mixtures are a linear combination of compositions bounded by the epilimnion and hypolimnion compositions on the ends (*dashed black line*), here representing the composition of peak density (as would be expected prior to the first overturn event) and epilimnion composition upon overturn. The resulting mole fraction of dissolved nitrogen may be lower than equilibrium (*blue shaded region*), resulting in atmospheric nitrogen dissolving into the mixed lake. Conversely, the combined composition may be supersaturated in nitrogen (*red shaded region*), resulting in an unstable composition that will exsolve nitrogen. Such nitrogen exsolution may be very rapid, and trigger bubble formation (e.g., *Farnsworth et al. 2020*).

Although the shape of the methane-ethane-nitrogen liquidus curve is poorly understood under Titan conditions, methane-rich mixtures can remain liquid below ~80K (*Hanley et al. 2017*), while ethane-rich mixtures begin to freeze around ~89K (*Richardson et al. 2019; Farnsworth et al. 2017; 2020*). Thus, evaporative methane-loss will eventually evolve the epilimnion across the liquidus curve, causing ethane ice to precipitate out of solution and thus stabilizing the epilimnion's composition.

2.3 *Experimental Verification of Stratification*

We also carried out laboratory studies to investigate the interactions between methane, ethane, and nitrogen, conducting experiments at the Northern Arizona University (NAU) Astrophysical Materials Laboratory. A detailed description of this facility is available in Tegler et al. (*2012*) and Grundy et al. (*2011*). We used a similar setup to Roe and Grundy (*2012*) and Protopapa et al. (*2015*). We prepared our samples by first mixing methane and ethane gases at the desired ratio in a two-liter mixing chamber at room temperature. We then cooled that sample using a helium compressor to cool the cell to the desired temperature. Once the cell was at the desired temperature, we obtained a background spectrum of the empty cell with our Raman spectrometer (Kaiser Optical Systems model Rxn1 using a 785 nm laser). We then opened the line between the cell and the mixing chamber, allowing the sample to condense into the cell. We allowed the sample to reach equilibrium. We then slowly added nitrogen gas until we reached the desired pressure, and waited until the temperature and pressure stabilized to ensure the sample reached equilibrium. We measured the composition of layers in the liquid with Raman spectroscopy using a model in which the molar fraction of a species in the liquid is linearly proportional to the band area in the spectrum. Photographs and videos were taken throughout the experiment.

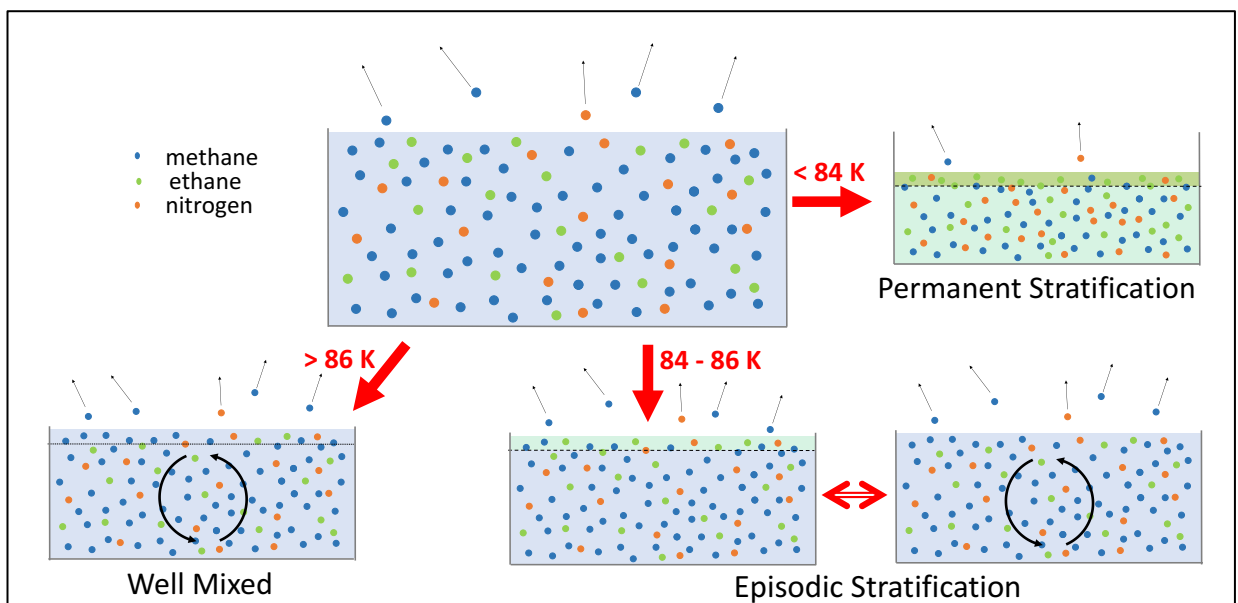

**Figure 4:** *Modes of Stratification.* The temperature of the lake determines the mode of stratification. Above 86K, a liquid methane–ethane–nitrogen mixture on Titan will remain well mixed, preventing stratification. Between 84 and 86K, the mixture becomes polymictic, undergoing cycles of stratification and overturn, until the bulk composition is poorer in methane than the local density minimum and becomes holomictic. Below 84K, the mixture becomes meromictic, eventually forming a nearly methane-free epilimnion.

For this specific experiment, we used a sample of 50% methane and 50% ethane at a fixed temperature, stepping from 84K to 87K in 1K increments. We added a total of 53.3 kPa (400 Torr) of hydrocarbons: 26.7 kPa (200 Torr) each of methane and ethane. At 83K, nitrogen was added to reach a pressure at or slightly above 1.47 bar (1100 Torr), until equilibrium was achieved. We then increased the temperature in 1K increments, consequentially increasing pressure. At each temperature, we agitated the cell to ensure the sample was homogeneous. We then reduced the pressure to 145 kPa (1090 Torr; Titan's surface pressure), which resulted in nitrogen and methane exsolving from the sample's surface. The sample stratified into two separate layers separated by a visible chemocline (Figure 5). We performed Raman spectroscopy at 1 mm intervals in each layer. Measurements in a layer were of nearly uniform composition. As a result, we computed the average and standard deviation of the composition in each layer.

As the sample equilibrated, the separation between the layers weakened, demonstrating that nitrogen and methane exsolution was required for layer formation. We then agitated the sample to check for liquid-liquid-vapor equilibrium (*e.g., Hanley et al.* 2016; *Cordier et al.* 2017). Upon agitation, the layers mixed, demonstrating that the layering was the result of exsolution-driven stratification, rather than the formation of two immiscible liquid phases (as would be present in a liquid-liquid-vapor equilibrium).

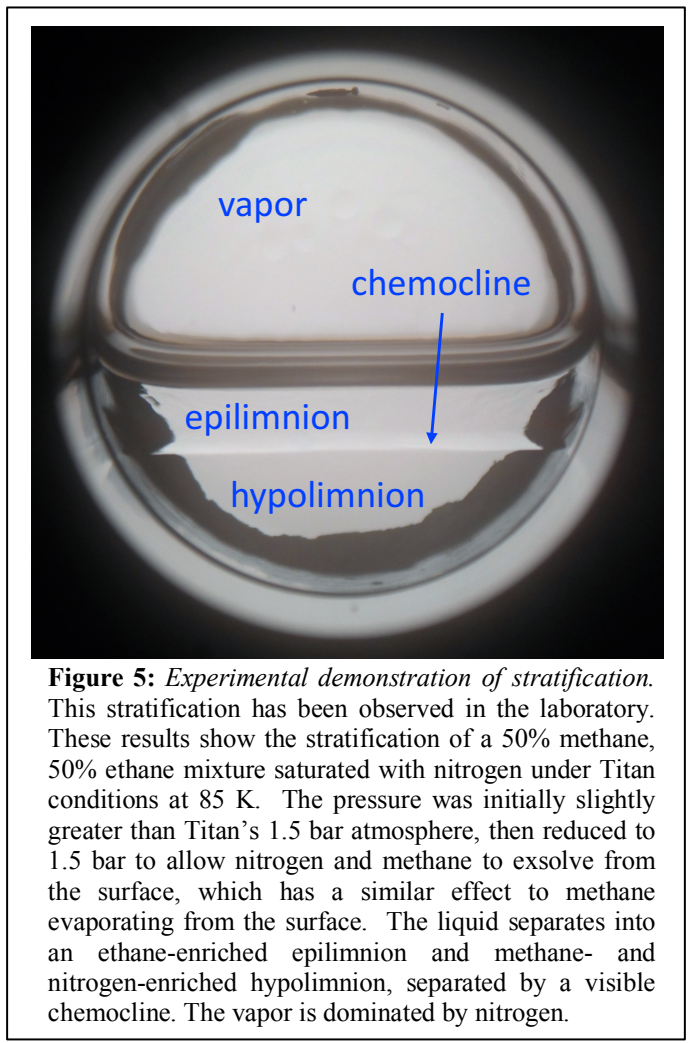

*Experimental demonstration of stratification.* This stratification has been observed in the laboratory. These results show the stratification of a 50% methane, 50% ethane mixture saturated with nitrogen under Titan conditions at 85 K. The pressure was initially slightly greater than Titan's 1.5 bar atmosphere, then reduced to 1.5 bar to allow nitrogen and methane to exsolve from the surface, which has a similar effect to methane evaporating from the surface. The liquid separates into an ethane-enriched epilimnion and methane- and nitrogen-enriched hypolimnion, separated by a visible chemocline. The vapor is dominated by nitrogen.

We carried out an additional experiment to determine if layering can form via temperature-driven exsolution. In this experiment, we mixed a liquid sample of 64% methane and 36% ethane at 90K. Then, we added nitrogen until we reached a pressure of 1090 torr (i.e., Titan's surface pressure). We reduced the temperature to 82 K and then raised the temperature to 83K, keeping the pressure at 1090 torr, which resulted in nitrogen and methane exsolving from the sample surface. The liquid again separated into two layers, and we took spectra of each layer.

In Table 1, we tabulate our nitrogen and alkane compositional results. The first two rows are from the temperature-driven exsolution experiments, and the following eight rows are from the pressure-driven exsolution experiments. For each experiment, we provide $X_{N2}$, $X_{CH4}/X_{C2H6}$, and $X_{CH4}/(X_{CH4}+X_{C2H6})$ for the epilimnion (upper layer) and hypolimnion (lower layer). Notice that in all five experiments $X_{N2}$, $X_{CH4}/X_{C2H6}$ and $X_{CH4}/(X_{CH4}+X_{C2H6})$ are smaller in the epilimnion (upper layer) than the hypolimnion (lower layer). In other words, the upper layer is richer in ethane and the lower layer is richer in methane and nitrogen as predicted by our numerical model.

Lastly, we used these experimental results to check the accuracy of the GERG-2008 equation of state in its "Extended Range of Validity" (see section 2.1). We focused on the composition of the epilimnia (upper layer), since the upper layer is equilibrated with the vapor phase (indeed, this is why the upper layer grows from the surface boundary). We input the temperature and composition of the upper layer into REFPROP (which implements GERG-2008), and output the expected pressure. We then compare this expected pressure to the measured pressure, and find that GERG-2008 produces errors no larger than 6.48% for our experimental results (see Table 1). This is consistent with GERG-2008's expected error (*Kunz & Wagner 2012*) and previous experimental results (*Richardson et al. 2018*), which confirm that GERG-2008 and REFPROP are sufficiently accurate for our purposes.

3. DISCUSSION

Stevenson and Potter (*1986*) first suggested that a methane–nitrogen mixture condensing from Titan's atmosphere would float atop, and thus stratify, any hypothetical (*Lunine et al. 1983*) global ethane–nitrogen ocean. Tokano (*2009*) later found that stable liquid lakes on Titan may seasonally stratify from summer heating, annually overturn and mix during the colder part of the year. However anomalous geothermal heating or evaporative cooling could prevent Titan's lakes from stratifying solely through thermal density changes (*Tokano 2009*). Recent computational (*Cordier et al. 2017*) and experimental (*Hanley et al. 2016*) studies find that pressures deep in Titan's largest seas are potentially high enough to separate methane–ethane–nitrogen liquid mixtures into two distinct liquid phases, separated by an interface similar to the liquid-vapor interface present at the surface of the sea. However, this cannot occur within Titan's ephemeral lakes, which are too shallow for such a phase separation. Instead, the layers of Titan's ephemeral lakes remain fully miscible, but nevertheless separated due to significant density differences.

| T (K) | P (Torr) | $X_{CH4}/(X_{CH4}+X_{C2H6})$ (Initial) | Layer | $X_{N2}$ (Layers) | $X_{CH4}/(X_{CH4}+X_{C2H6})$ (Layers) | $P_{REFPROP}$ (pred.) (Torr) | Model Error % |
|---|---|---|---|---|---|---|---|
| 83 | 1090 | 0.64 | U | 0.455 ± 0.021 | 0.606 ± 0.013 | 1072 | 1.65 |
| 83 | 1090 | 0.64 | L | 0.560 ± 0.006 | 0.635 ± 0.007 | … | … |
| 84 | 1093 | 0.50 | U | 0.197 ± 0.006 | 0.487 ± 0.022 | 1110 | 1.02 |
| 84 | 1093 | 0.50 | L | 0.251 ± 0.003 | 0.503 ± 0.004 | … | … |
| 85 | 1090 | 0.50 | U | 0.161 ± 0.005 | 0.477 ± 0.020 | 1123 | 3.03 |
| 85 | 1090 | 0.50 | L | 0.191 ± 0.004 | 0.501 ± 0.003 | … | … |
| 86 | 1090 | 0.50 | U | 0.138 ± 0.004 | 0.477 ± 0.025 | 1123 | 3.03 |
| 86 | 1090 | 0.50 | L | 0.162 ± 0.005 | 0.498 ± 0.003 | … | … |
| 87 | 1090 | 0.50 | U | 0.124 ± 0.010 | 0.477 ± 0.026 | 1140 | 6.48 |
| 87 | 1090 | 0.50 | L | 0.135 ± 0.003 | 0.498 ± 0.002 | … | … |

**Table 1:** *Data from experimental effort.* This table reports our experimental data for five experimental runs. The first two rows (83K) present our temperature-driven exsolution experiment results, while the next eight rows (84K – 87K) present our pressure-driven exsolution experiment results. The lower layer (hypolimnion) is enriched in methane for all five experiments. Uncertainties are the standard deviations of measurements in a layer. An ethane enriched upper layer and a methane-nitrogen enriched lower layer confirms our numerical model in this temperature range of interest. We also compute the expected pressure from the temperature and composition of the sample's upper layer to validate our TITANPOOL code, and check that errors are small (on the order of a few percent).

In contrast to these schemes, we find that Titan's ephemeral lakes chemically stratify via evaporation-induced compositional changes. Challenging prevailing thought, we find that nitrogen's high solubility in methane-rich mixtures significantly increases its density beyond that of an ethane-rich liquid, allowing lakes to evolve *ethane-rich* epilimnia and *methane-rich* hypolimnia. This chemical stratification may generate discrepancies in composition derived from RADAR-derived dielectric constant measurements of the lake (which probes the bulk lake composition) and spectroscopic measurements (which only samples the surface layer). As such, stratification could be detected by identifying significant deviations in composition between spectroscopic (e.g., surface) and RADAR (e.g., bulk lake) observations. However, spectroscopic observations need to account for atmospheric methane and nitrogen; at present, insufficient laboratory data exist at present to derive compositional constraints from VIMS observations.

Such stratification may explain sudden increases in surface albedo observed to follow Titan's rain storms (*Barnes et al. 2013; Soderblom et al. 2016*). As evaporation cools and stratifies the resulting ephemeral lakes, their epilimnia become ethane-rich and may cross the liquidus curve. Above ~83 K, any ethane ice that precipitates out of solution would be denser (*Klimenko et al. 2008*) than the surrounding liquid and settle to the lake bed, where it would remain stably frozen. Evaporation of the remaining lake liquids would eventually expose this ice deposit, which would appear as a very rapid brightening of a dark lake, consistent with observations (*Barnes et al. 2013; Soderblom et al. 2016*). Titan's polar "phantom lakes" may also stratify, however it is unclear if these lakes are disappearing due to evaporation or infiltration (*MacKenzie et al. 2019*). Infiltration would not lead to stratification unless the infiltration rate varies significantly between the lakes' constituent species (but would be largely temperature independent).

Stratification can spontaneously occur without methane evaporation, if inflowing liquids arrive out of thermochemical equilibrium with the lake. The resulting density differences may be significant, with new material either floating into an epilimnion, or sinking to the bottom in a submarine plume. This process is density driven and qualitatively no different than evaporation-induced stratification. Methane-rich layers will always form the epilimnion above 86 K due to their lower density; below 86 K, such lakes are susceptible to the stratification cycles and dynamics described above. Nevertheless, this process requires runoff to flow into an existing liquid body. Since the observed ephemeral ponds (*Barnes et al. 2013; Soderblom et al. 2016*) fill otherwise dry playa, they are unlikely to stratify in this manner.

4. CONCLUSIONS

We constructed and used the TITANPOOL numerical code to explore the effects of nonideal interactions between methane, ethane, and nitrogen liquid mixtures on Titan's surface. We find that such interactions lead to strong temperature and composition dependences on dissolution of atmospheric nitrogen, which significantly affects the liquids' densities. Disequilibrium liquid temperature resulting from air-sea interactions (*Rafkin & Soto 2019*) or fresh precipitation (*Graves et al. 2008*) could, thus, produce methane-rich mixtures that are denser than more ethane-rich liquids. Methane evaporation can therefore cause shallow lakes to stratify; the evaporative dynamics of the lake depend on temperature. We verified this methane evaporation-induced stratification in the laboratory. Such dynamics could influence the dynamic behavior of ephemeral ponds that form following rainstorms (*Barnes et al. 2013; Soderblom et al. 2016*) and the north polar "phantom" lakes (*MacKenzie et al. 2019*), and influence limnological dynamics near deltas (where rivers flow into lakes).

5. ACKNOWLEGEMENTS

We wish to thank Jason Barnes and an anonymous reviewer, whose comments greatly improved this manuscript. We graciously acknowledge Dr. Wolfgang Wagner in helping us understand the accuracy and validity of the GERG-2008 model below 90K. We also acknowledge Dr. Chris Glein, whose thoughtful discussions helped us understand how best to model hydrocarbon–nitrogen mixtures and methane evaporation. J.K.S., J.M.S., K.F., and V.C. were supported in part by NASA Cassini Data Analysis grant NNX15AL48G. J.K.S., J.M.S., and A.S. were also supported in part by NASA Cassini Data Analysis grant 80NSSC18K0967. J.H., J.G., W.G., L.G., S.T., and A.E. were supported in part by NASA Solar System Workings grant #80NSSC18K0203, NSF grant number AST-1461200 and a grant from the John and Maureen Hendricks Charitable Foundation.